\documentclass[twocolumn,prb,showpacs,aps,superscriptaddress]{revtex4-1}
\usepackage{amsmath}
\usepackage{amssymb}
\usepackage{amsfonts}
\usepackage{bm}
\usepackage{amssymb}
\usepackage[dvips]{graphicx}
\usepackage{dcolumn}
\usepackage{txfonts}
\usepackage{makeidx}
\usepackage{color}
\usepackage{mathtools}
\usepackage{threeparttable}
\usepackage[linkcolor=blue,anchorcolor=black,citecolor=blue]{hyperref}
\usepackage{amsmath}
\usepackage{amssymb}
\usepackage{multirow}
\usepackage{graphicx}
\usepackage{dcolumn}
\usepackage{bm}
\usepackage{esint}
\usepackage{color}

\begin{document}

\title{Phase transition of light in circuit QED lattices coupled to
nitrogen-vacancy centers in diamond}
\author{Jia-Bin You}
\email{jiabinyou@gmail.com}
\affiliation{Center for Quantum Technologies and Physics Department, National
University of Singapore, 3 Science Drive 2, 117543, Singapore}
\author{W. L. Yang}
\email{ywl@wipm.ac.cn}
\affiliation{Center for Quantum Technologies and Physics Department, National
University of Singapore, 3 Science Drive 2, 117543, Singapore}
\affiliation{State Key Laboratory of Magnetic Resonance and Atomic
and Molecular Physics, Wuhan Institute of Physics and Mathematics,
Chinese Academy of Sciences, Wuhan 430071, China}
\author{Zhen-Yu Xu}
\affiliation{School of Physical Science and Technology, Soochow University, Suzhou
215006, China}
\author{A. H. Chan}
\affiliation{Center for Quantum Technologies and Physics Department, National
University of Singapore, 3 Science Drive 2, 117543, Singapore}
\author{C. H. Oh}
\email{phyohch@nus.edu.sg}
\affiliation{Center for Quantum Technologies and Physics Department, National
University of Singapore, 3 Science Drive 2, 117543, Singapore}

\begin{abstract}
We propose a hybrid quantum architecture for engineering a photonic Mott
insulator-superfluid phase transition in a $2D$ square lattice of
superconducting transmission line resonator (TLR) coupled to a single
nitrogen-vacancy (NV) center encircled by a persistent current qubit. The
localization-delocalization transition results from the interplay between
the on-site repulsion and the nonlocal tunneling. The phase boundary
in the case of photon hopping with real-valued and complex-valued amplitudes can be obtained using the mean-field approach. Also, the quantum jump
technique is employed to describe the phase diagram when the dissipative
effects are considered. The unique feature of our architecture is the good
tunability of effective on-site repulsion and photon-hopping rate, and
the local statistical property of TLRs which can be analyzed readily using present microwave techniques. Our work opens new perspectives in quantum
simulation of condensed-matter and many-body physics using a hybrid spin
circuit QED system. The experimental challenges are
realizable using current available technologies.
\end{abstract}

\maketitle

\section{Introduction}

The microscopic properties of strongly correlated
many-particle systems emerging in solid-state physics are in general very
hard to access experimentally \cite{ref1,ref2}. So how to simulate the
properties of condensed-matter models using nontraditional controllable
systems is highly desirable. Recently, the investigation of quantum
simulation in the photon-based many-body physics has received much attention
in different systems \cite{ref1,ref2,ref3,ref4}. Especially, there has been
a great interest in mimicking quantum phase transition (QPT) of light with
scalable coupled resonator array in the context of cavity/circuit quantum
electrodynamics (QED) \cite{Lep,Houck,Cir1,Cir2}, which provides a
convenient controllable platform for studying the strongly correlated states
of light via photonic processes. On the other hand, the artificially
engineered hybrid devices can permit measurement access with unique
experimental control \cite{Fazio,Lew}; and it is intriguing to employ a
well-controllable quantum system with a tunable Hamiltonian to simulate the
physics of another system of interest. This paradigm has promoted many
experimental/theoretical proposals on probing the light phase and opened
various possibilities for the simulation of many-body physics.
\begin{figure}[tbp]
\includegraphics[width=6cm]{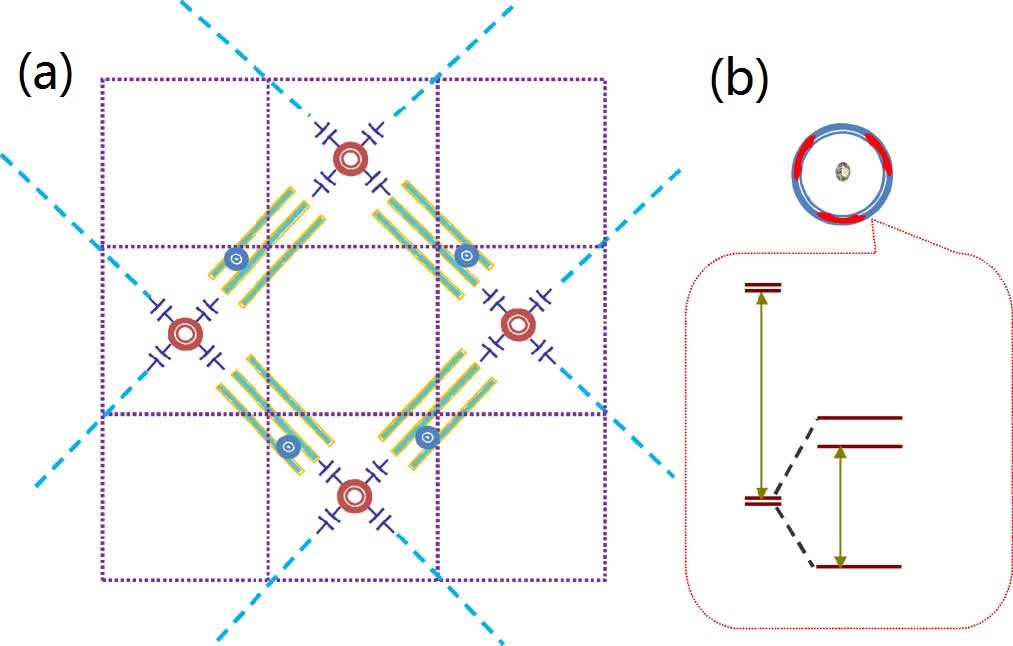}
\caption{(Color online) (a) Schematic circuit for the resonator lattice,
where each TLR is coupled to a single NV center encircled by a PCQ, and the
circles denote the central coupler. (b) The subsystem consisting of NV and
PCQ, where NV is at the center of the loop. The PCQ is made up of three
Josephson junctions, and it couples to the NV via the magnetic field at the
center of the loop generated by the persistent currents in the loop. The
energy diagram of the NV is shown in the red box.}
\end{figure}

In this work, we develop an optical system for engineering the strongly
correlated effects of light in a hybrid solid-state system. We consider a $2D$ square lattice of coupled TLRs \cite{SC}, where each TLR is coupled to a
single NV \cite{NV1,QP} encircled by a persistent current qubit (PCQ). We
show that the competition between the NV-PCQ-TLR interaction and the
nonlocal hopping induces the photonic localization-delocalization transition. Subsequently the Mott insulator (MI) phase and the superfluid (SF) phase can appear in a controllable way. The phase boundary in the case of photon
hopping with real/complex-valued amplitudes can be obtained using the
mean-field approach. Also, the quantum jump technique is employed to
describe the phase diagram when the dissipation is considered. Finally, the
possibility of observation of the QPT is discussed by employing
experimentally accessible parameters.

In our architecture, one can tune independently the on-site emitter-field
interaction and the nonlocal photonic hopping between adjacent TLRs. This permits us to systematically study the localization-delocalization
transition of light in a complete parameter space. The main motivation for
building such a hybrid system is to combine several advantages: \textit{in situ} tunability of circuit elements \cite{Cir3}, spectroscopic technology for state readout, peculiar characteristics of NV (e.g., individual addressing and long coherence time at room-temperature \cite{NV2}), and scalability of TLR arrays \cite{Houck,Koch,Nu1,Nu2}. Recently, D. Underwood \textit{et al} experimentally fabricated 25 arrays of TLRs and demonstrated the feasibility of quantum simulation in circuit QED system \cite{Und}. E. Lucero \textit{et al } experimentally characterized a complex circuit composed of four phase qubits and five TLRs to realize intricate quantum algorithms \cite{Lu}. The
progress renders the TLR lattice a good platform for studying
condensed-matter physics with photons and makes our scheme to be more
practical.
\begin{figure}[tbp]
\includegraphics[width=6.0 cm, bb=81pt 442pt 502pt
767pt]{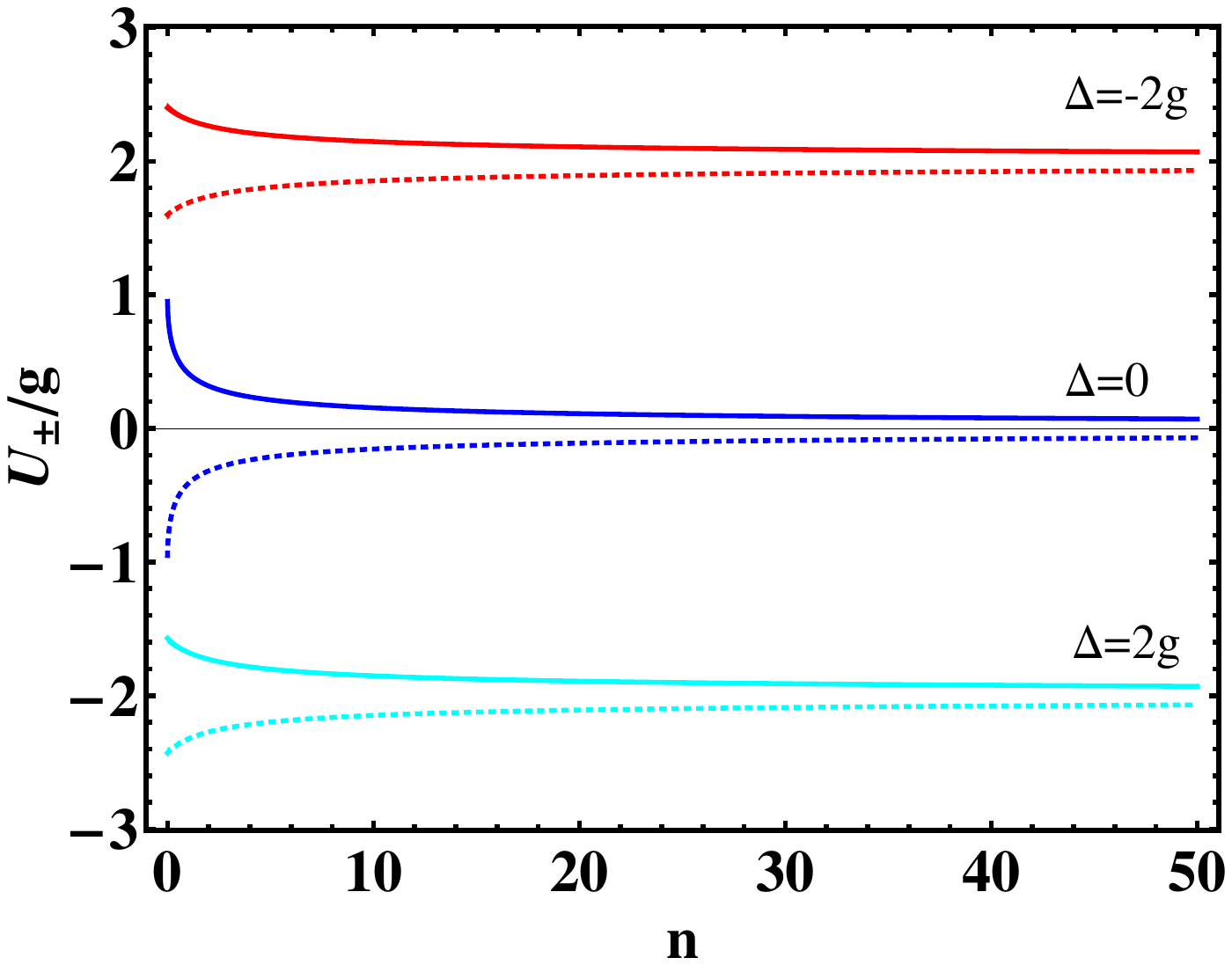}
\caption{(Color online) The dependence of effective on-site repulsion $%
U_{\pm }$ on the photon number $n$ under the different detunings $\Delta $,
where the solid (dotted) line denotes $U_{+}$ $(U_{-})$.}
\end{figure}

\section{Model}

As illustrated in Fig. 1, we consider a $2D$ lattice of
coupled TLRs, where the basic unit consists of a TLR coupled to a single NV
encircled by a PCQ, which acts as an interconnect to greatly magnify the
NV-TLR coupling by several orders of magnitude, compared with the direct
NV-TLR coupling (far below the linewidth of TLR with dozens of $kHz$)
resulting from the vacuum fluctuations of the photons \cite{Vac,Sys}. The TLR is made of a superconductor line interrupted by two capacitors at its ends. In the microwave domain, it can be treated as a quantum LC harmonic oscillator, $H^{TLR}=\omega _{r}(a^{+}a+1/2)$ ($\hbar =1$), where $\omega_{r}=\sqrt{1/L_{r}C_{r}}$ is the corresponding eigenfrequency with inductance $L_{r}$ and capacitance $C_{r}$.
The PCQ located at an antinode of TLR's magnetic field is formed by a superconducting loop interrupted by three Josephson junctions \cite{Pcq}. When the loop is biased by half a magnetic flux quantum, the device is an effective two-level qubit made up of two countercirculating persistent currents with the Hamiltonian $H^{PCQ}=\frac{\omega _{0}}{2}\sigma _{z}$. The NV can be modeled as a three-level system in the triplet
ground-state subspace consisting of $\left\vert ^{3}A,m_{s}=0\right\rangle$
and $\left\vert ^{3}A,m_{s}=\pm 1\right\rangle$. The Hamiltonian is $H^{NV}=\gamma _{e}B_{z}S_{z}+D(S_{z}^{2}-2/3)$, where $\gamma _{e}$ is the electronic gyromagnetic ratio, $D/2\pi \sim 2.87GHz$ is the zero-field
splitting, $B_{z}$ is a perpendicular magnetic field at the NV and $S_{z}$ is the spin-1-$z$ operator.

The PCQ magnetically couples to TLR via mutual inductance, $H^{T-P}=-\hat{\mu}\cdot\hat{B}$, where $\hat{\mu}$ is the magnetic dipole of PCQ induced by the persistent circulating currents and $\hat{B}$ is the magnetic field at PCQ induced by the current in the central
conductor of TLR. When $\omega_{r}\sim\omega_{0}$, we have $H^{T-P}=g(a^{+}\sigma ^{-}+a\sigma ^{+})$ after rotating wave approximation, where $g=(I_{p}\mu _{0}r^{2}/d)\sqrt{\omega _{r}/2L_{r}}$, $r$ $(I_{p})$ is the radius (persistent circulating current) of the PCQ loop, and $d$ is the distance between PCQ and central conductor of TLR.
The sizable changes of magnetic flux within the loop induced by $I_{p}$ presented in the PCQ lead to small shifts in the transition frequencies ($m_{s}\rightarrow \pm 1$) of NV \cite{Sys,Pcq}. Through this small change in magnetic field the PCQ can couple to the NV via Zeeman term, $H^{N-P}=\frac{1}{2}\eta\sigma_{z}S_{z}$, where $\eta=I_{p}\mu_{0}\gamma_{e}/r$.

The basic unit in our system is thus governed by the
Hamiltonian $H_{p}^{0}=H^{TLR}+H^{PCQ}+H^{NV}+H^{T-P}+H^{N-P}$. The photonic tunneling in our model can be realized by a central coupler \cite{expl1} which serves as individual tunable quantum transducers to transfer photonic states between adjacent TLRs. We have presented a new paradigm for $2D$ TLR lattice coupled to solid-state spins. We have shown that specially engineered resonator lattice provides a practical platform to couple both individual spin and superconducting qubit, and engineer their interactions in a way that surpasses the limitations of current technologies. This can provide new insights to many-body physics.

\section{Mott-superfluid transition}

We study the full Hamiltonian of the $2D$ square lattice by adding the on-site chemical potential and the
nonlocal microwave photon hopping between adjacent sites. The Hamiltonian is
given by
\begin{equation}
H=\sum_{p}H_{p}^{0}+\sum_{\left\langle p,q\right\rangle }k_{\left\langle
p,q\right\rangle }a_{p}^{+}a_{q}-\sum_{p}\mu _{p}N_{p},
\end{equation}
where $k_{\left\langle p,q\right\rangle }=$ $2Z_{0}C_{\left\langle
p,q\right\rangle }(\omega _{r}+\delta_{p})(\omega_{r}+\delta_{q})$ are
photonic tunneling rates between resonators $p$ and $q$, which are set by
the tunable mutual capacitance $C_{\left\langle p,q\right\rangle }$ between resonator ends with characteristic impedance $Z_{0}$ and frequency shift $\delta _{p}$ due to random disorder \cite{Und}. Since $\omega _{r}\gg\delta
_{p}$, one can assume that $k_{\left\langle p,q\right\rangle}=k=2Z_{0}C\omega_{r}^{2}$ without disorder for nearest-neighbor
resonators, and $k_{\left\langle p,q\right\rangle }$ $=0$ for other
resonator pairs. $\mu _{p}$ is the chemical potential at the $p$-th site.
The conserved quantity in our system is the total number of excitations $%
N_{p}=a_{p}^{+}a_{p}+\sigma _{p}^{+}\sigma _{p}^{-}+\frac{1}{2}%
S_{p}^{+}S_{p}^{-}$ with $S^{i}$ $(\sigma^{i})$ the spin-$1$ (-$1/2$)
operators $(i=x,y,\pm )$.

The photon-dependent eigenstates of the Hamiltonian $H_{p}^{0}$ is dressed states $\left\vert m_{s},\pm ,n\right\rangle$ with the eigenvalues $E_{\left\vert m_{s},\pm ,n\right\rangle }=n\omega _{r}+(-\Delta +2n\Delta
\pm \sqrt{4ng^{2}+\Delta ^{2}\pm 2\Delta \eta +\eta ^{2}})/2+\chi (m_{s})$,
where $\chi (m_{s})=D(3\times 1^{m_{s}}-2)+m_{s}\gamma _{e}B_{z}$ is the
eigenenergy of $H^{NV}$. Here $\Delta =\omega _{r}-\omega _{0}$ is the
detuning, $n$ is the number of excitations in the resonator and $|\pm
\rangle =$ $(|e\rangle \pm |g\rangle )/\sqrt{2}$. In our case, the dynamics
is governed by the Jaynes-Cummings (JC) type of interaction, which enables
the interconversion of qubit excitations and photons, and provides the
effective on-site repulsion. Meanwhile, pairs of TLRs are coupled by the
two-site Hubbard model via one-photon hopping. The difference between the
Bose--Hubbard model (BHM) \cite{BHM} and our model is that the conserved
particles are the polaritons rather than the pure bosons in BHM, and the
effective on-site repulsion $U_{\pm }(n)=E_{\left\vert m_{s},\pm
,n+1\right\rangle }-E_{\left\vert m_{s},\pm ,n\right\rangle }$ decreases
with the growth of photon number, and $U_{\pm }(n)\rightarrow 0$ in the
limit of large $n$ and $\Delta =0$, as shown in Fig. 2, while it is a
constant in BHM.

\begin{figure}[tbp]
\includegraphics[width=7.5 cm, bb=4pt 1pt 461pt
528pt]{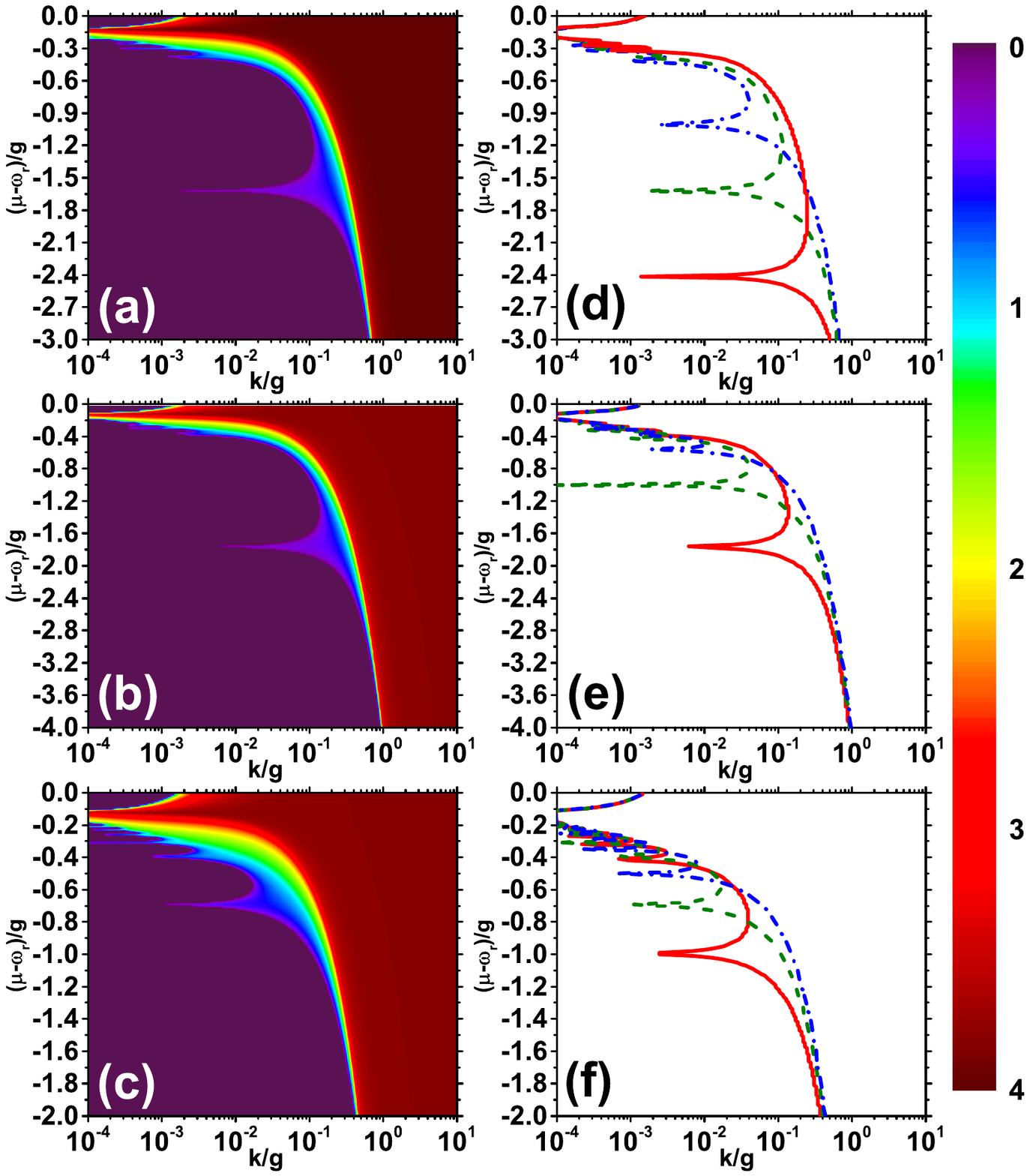}
\caption{(color online). The phase diagrams in $\protect\mu\sim k$ plane for
different sets of NV-PCQ coupling $\protect\eta$ and tunable magnetic fields
$B_{z}$ applied on NV, where $\protect\mu$ and $k$ are the chemical
potential and the photon hopping rate. The common parameters are $g=1$, $%
\protect\omega _{r}=200$, $D=100$ and $\protect\gamma _{e}=-10^{3}$. The
other parameters are set as (a) $\protect\eta =0.01$, $B_{z}=0.0005 T$ and $%
\Delta =g$, the phase boundaries are plotted in (d), where $\Delta =2g$
(solid), $g$ (dashed) and $0$ (dot-dashed); (b) $\protect\eta =1.2$, $%
B_{z}=-0.3$ and $\Delta =0$, the phase boundaries are plotted in (e), where $%
B_{z}=-0.3 T$ (solid), $0.0005 T$ (dashed) and $0.3 T$ (dot-dashed); (c) $%
\protect\eta =0.75$, $B_{z}=0.3 T$ and $\Delta =0$, the phase boundaries are
plotted in (f), where $\protect\eta=0.01$ (solid), $\protect\eta =0.75$
(dashed) and $\protect\eta=1.5$ (dot-dashed).}
\label{qpt1}
\end{figure}
The phase diagrams can be distinguished using the corresponding order
parameters.\textbf{\ }Here we choose the SF order parameter $\psi =\langle
a_{p}\rangle $ (set to be real) to differentiate between insulator-like and
SF-like states. Using the mean-field theory \cite{MF}\, we decouple the
hopping term as $a_{p}^{+}a_{q}=$ $\left\langle a_{p}^{+}\right\rangle
a_{q}+a_{p}^{+}\left\langle a_{q}\right\rangle -\left\langle
a_{p}^{+}\right\rangle \left\langle a_{q}\right\rangle $, the resulting
mean-field Hamiltonian can then be written as a sum over single sites,
\begin{equation}
H^{MF}=\sum_{p}[H_{p}^{0}-zk\psi (a_{p}^{+}+a_{p})+zk\psi ^{2}-\mu
_{p}N_{p}],
\end{equation}
where $z=4$ is the number of nearest neighbours. Noting that $%
[H^{MF},S_{z}]=0$, therefore, we can treat $S_{z}$ in the mean-field
Hamiltonian as a \textit{c-}number and $S_{z}=\pm 1,0$. Minimizing the
ground state energy of the Hamiltonian $H^{MF}$ with respect to $\psi $ for
different values of $\mu $ and $k$, we obtain the mean field phase
diagram/boundary in the $(\mu,k)$ plane when $\eta $ varies from the weak
coupling regime $(\eta \ll g)$ to the strong coupling regime $(\eta >g)$
under the resonant/detuning case. The features of Fig. 3 are rich. The regime where $\psi =0$ corresponds to the stable and incompressible MI lobes characterized by a fixed number of excitations at per site with no variance. In each MI lobe, due to the nonlinearity and anharmonicity in the
spectrum originating from the photon blockade effect \cite{Blo}, the strong
emitter-field interaction leads to an effective large polariton-polariton
repulsion which freezes out hopping and localizes polaritons at individual
lattice sites. By contrast, strong hopping favours delocalization and
condensation of the particles into the zero-momentum state, namely, $\psi
\neq 0$ indicates a SF compressible phase with the stable ground state at
each site corresponding to a coherent state of excitations.

Analogous to the BHM,\emph{\ }the physical picture behind is that the MI-SF
phase transition results from the interplay between polariton delocalization
and on-site repulsive interaction. Therefore, the phase boundary primarily
depends on the ratio of the photon-hopping rate to the on-site repulsion
rate. When the on-site repulsion dominates over hopping, the system should
be in the MI phase, otherwise the system will be in the SF phase. From the
expression of the parameter $\eta $ and $g$, we can find that reduction of
the size of the PCQ loop will increase $\eta $ but decrease $g$, and the
adjustment of the distance $d$ only affects TLR-PCQ interaction.
Furthermore, the detuning $\Delta $ is also tunable by varying the magnetic
field applied on NV. In Fig. 3, one can find that the size of the MI lobes
varies with $\Delta $, with the largest Mott lobes found on resonance.

Further insight to the transition can be gained when the photon hopping with
complex-valued amplitude exists in Eq. (1), where the hopping process
becomes $-\sum\nolimits_{\left\langle p,q\right\rangle }k_{\left\langle
p,q\right\rangle }e^{i\phi _{\left\langle p,q\right\rangle }}a_{p}^{+}a_{q}$
with $\phi _{\left\langle p,q\right\rangle }=-\phi _{\left\langle
q,p\right\rangle }$ and we set $k_{\left\langle p,q\right\rangle }=k\text{.}$
We emphasize that this process is possible if the intermediate coupling
elements are used to break time-reversal symmetry \cite{Houck,Koch,Per}.
The parameter $k_{\left\langle p,q\right\rangle }e^{i\phi
_{\left\langle p,q\right\rangle }}$ provides a new regime in the dynamical
evolution of the system. The sum of the tunneling phases along a closed loop
surrounding the plaquette is $2(\phi _{p+1,q}+\phi _{p,q+1}-\phi _{p,q}-\phi
_{p+1,q+1})=2\pi \alpha $, which is actually the flux quanta per plaquette.
Assuming that $\alpha $s are all equal, the total Hamiltonian under
mean-field approximation reads
\begin{eqnarray}
H_{\alpha }^{MF} &=&\sum\nolimits_{p}[H_{p}^{0}-zk\psi \cos (2\pi \alpha
)(a_{p}^{+}+a_{p})  \notag \\
&&+zk\psi ^{2}\cos (2\pi \alpha )-\mu _{p}N_{p}].
\end{eqnarray}%
The results are exhibited in Fig. 4, we find that the boundary line
gradually shifts to the right as $\alpha $\ enhances in the interval $[0,1/4]
$. Because of the spatial variation of tunneling phase, the wave function of
a polariton from one lattice site to another acquires a nontrivial phase (%
\textit{Aharonov-Bohm} phase) \cite{AB}, which actually reduces the
effective hopping rates.
\begin{figure}[tbp]
\includegraphics[width=10 cm, bb=1pt 2pt 671pt
276pt]{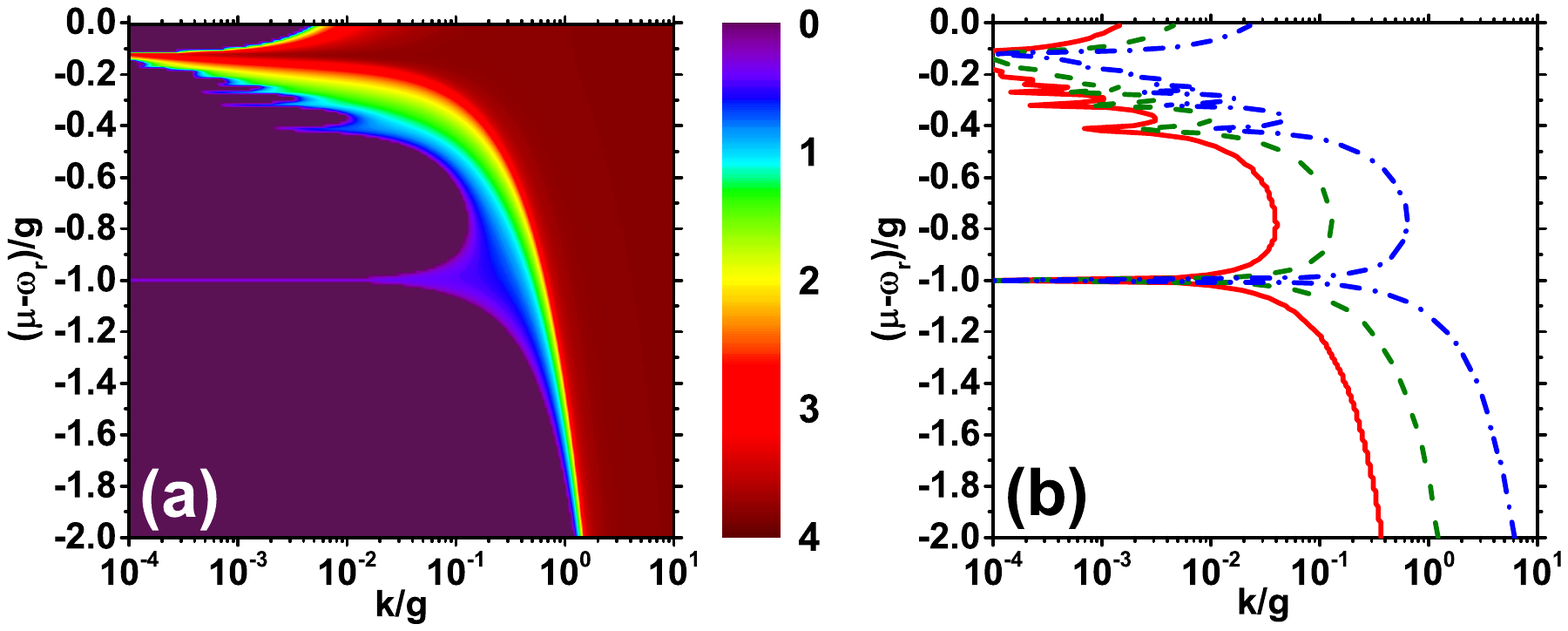}
\caption{(color online). (a) The order parameter $\protect\psi $ in $\protect%
\mu \sim k$ plane. The parameters are $g=1$, $\protect\omega _{r}=200$, $%
D=100$, $\protect\gamma _{e}=-10^{3}$, $\protect\eta =0.01$, $B_{z}=0.0005T$%
, $\protect\alpha =0.2$, and $\Delta =0$. The corresponding phase boundaries
are plotted in (b), where the solid, dashed, and dot-dashed line denote $%
\protect\alpha =0$, $0.2$, and $0.24$, respectively. }
\label{qpt2}
\end{figure}

\section{Dissipative effects}

Generally, nonequilibrium processes such as
dissipative effect, are crucial in solid-state devices. We show that the
signature of the localization-delocalization transition remains even in the
presence of the engineered dissipation by the quantum trajectory method \cite%
{QJ}. The non-Hermitian Hamiltonian is formulated as
\begin{equation}
H_{de}=H^{MF}-\frac{i\Gamma }{2}\sum\nolimits_{p}\sigma _{p}^{+}\sigma
_{p}^{-}-\frac{i\kappa }{2}\sum\nolimits_{p}a_{p}^{+}a_{p},
\end{equation}%
where $\kappa =4Z_{0}^{2}C_{out}^{2}\omega _{r}^{3}$ is the decay rate of
TLR, and $\Gamma $ is the decay rate from the effective excited state $%
|e\rangle $ of PCQ. In our case, the dissipative effects result from the unavoidable interaction between the PCQ/TLR and the corresponding Markovian
environment, for example, the interaction between the output of the TLR and the
corresponding vacuum field will result in a photonic escape rate with $%
\kappa $ to the continuum. Here the dissipative effects of NV are
negligible, compared with $\kappa $ and $\Gamma $. The phase diagrams under dissipative effects are displayed in Fig. 5. Once the hopping rate is
increased beyond a critical value, the system is expected to undergo a
non-equilibrium QPT from a MI phase, where the initial photon population is
self-trapped, to a SF phase with the dynamical photon population imbalance
coherently oscillating between pairs of TLRs \cite{Houck}. Furthermore,
another obvious feature is that the size of MI phase becomes larger as the
growth of dissipative rates. Note that the effective nonlinearity at per
site becomes stronger at lower exciton numbers, which implies that the
dissipative effect (inducing the decrease of the exciton numbers) favours
the MI phase. As a result, the dissipation results in the dynamical
switching from SF phase to MI phase and causes the increment of the size
of MI phase.
\begin{figure}[tbp]
\includegraphics[width=7.5 cm, bb=0pt 1pt 621pt
461pt]{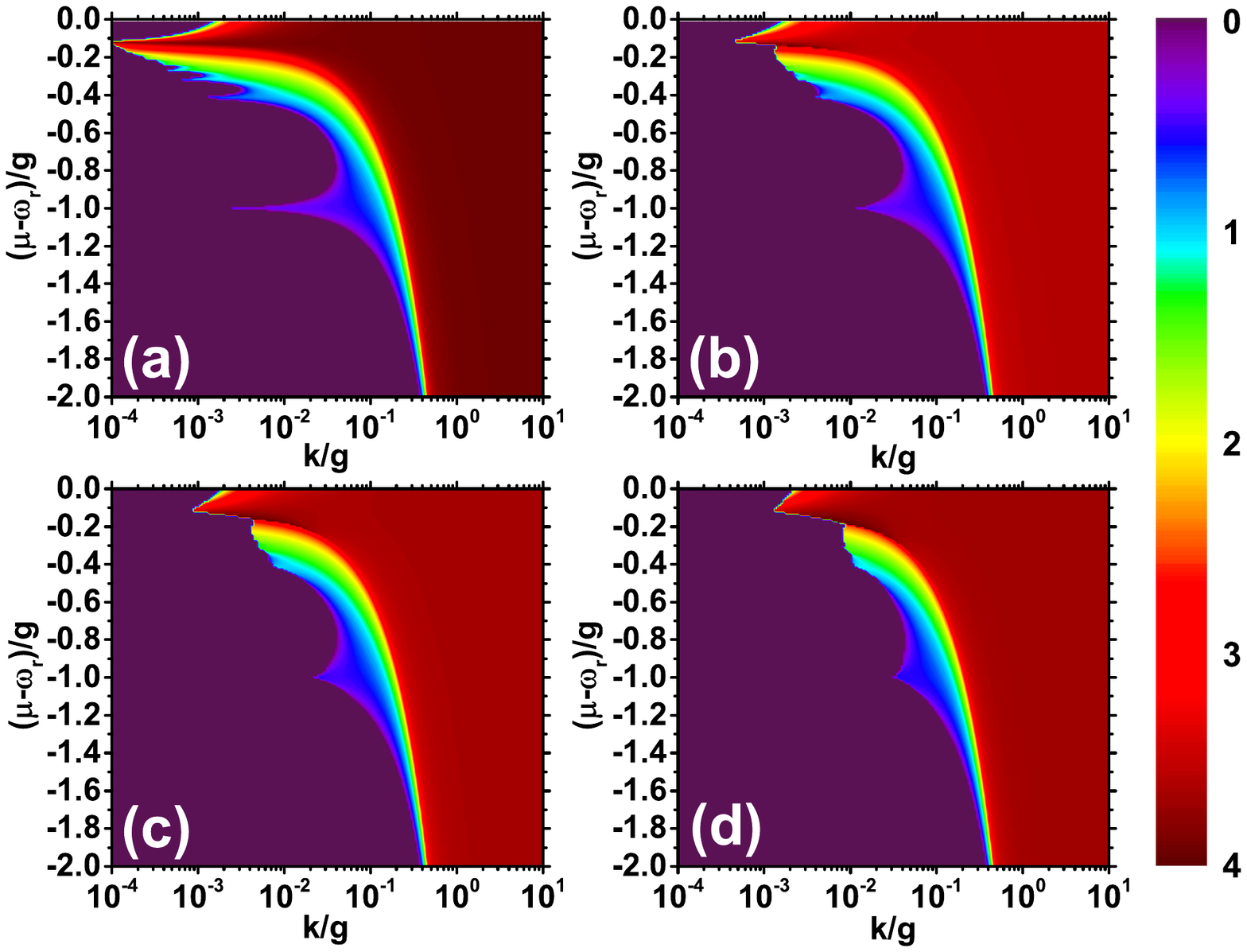}
\caption{(color online). The order parameter $\protect\psi $ in $\protect\mu %
\sim k$ plane under the different dissipative rates (a) $\Gamma =\protect%
\kappa =0.01$; (b) $\Gamma =$ $\protect\kappa =0.05$; (c) $\Gamma =$ $%
\protect\kappa =0.1$; (d) $\Gamma =\protect\kappa =0.15$. The other
parameters are $g=1$, $\protect\omega _{r}=200$, $D=100$, $\protect\gamma %
_{e}=-10^{3}$, $\protect\eta =0.01$, $B_{z}=0.0005 T$, and $\Delta =0$.}
\label{qpt3}
\end{figure}

\section{Experimental feasibility}

Firstly, we briefly stress the relevant
experimental progress. From theoretical standpoint, it is possible to fabricate large arrays to observe many-body physics of interacting polaritons since resonators and qubits can be made lithographically \cite{Ash}.
Actually, it is indeed feasible to couple over $200$ (or 1000) TLRs with
negligible disorders (on the order of a few parts in $10^{4}$) in a $2D$
lattice using a $32mm\times 32mm$ sample or a full two-inch wafer \cite%
{Houck}. Secondly, how to probe quantum many-body states of light is still
an open question in photonic quantum simulation \cite{Ma}. The previous
works \cite{ref2,Ang} suggested to measure the individual TLR through
mapping the excitations into the qubit followed by state-selective resonance
fluorescence spectrum, but a remaining technical challenge is the
realization of high-efficiency photon detectors. Alternatively, the local
statistical property of TLR can be analyzed readily using combined
techniques of photon-number-dependent qubit transition \cite{exper1,exper2}
and fast readout of the qubit state through a separate low-Q resonator mode
\cite{read}, for which the high-efficiency photon detectors are not
required. Experimentally, transmission and reflection measurements for
circuit QED arrays have been implemented successfully in small system with
one or two resonators \cite{Cir1,exper2}. Therefore, in order to distinguish
between different phases of the system, one can also experimentally probe
beyond transmission, such as two-tone spectroscopy and second-order
coherence function (photon statistics) to reveal additional information. The tunability of coupling strengths in our system enables one to measure these quantities relatively straightforwardly.

Finally, we survey the relevant experimental parameters. Given the
flexibility of circuit QED, we can access a wide range of tunable
experimental parameters for TLR-PCQ coupling strength $g$ and hopping rate $%
k_{\left\langle p,q\right\rangle }$. Taking $L_{r}=2$ $nH$, $\omega
_{r}/2\pi =6$ $GHz$, $I_{p}=600$ $nA$, and $r=0.2$ $\mu m$, we get $\eta
/2\pi \simeq 140$ $KHz$ and $g/2\pi \sim \lbrack 1.8,180]$ MHz when the
distance $d$ varies from $5$ $\mu m$ to $50$ $nm$. Furthermore, the hopping
rate $k_{\left\langle p,q\right\rangle }$ depends on the tunable mutual
capacitance $C_{\left\langle p,q\right\rangle }$ between resonator ends. In
Ref \cite{Und}, the authors measured devices with photon hopping rates $%
k/2\pi $ form $0.8$ MHz to $31$ MHz in resonators lattices. On the other
hand, the electron-spin relaxation time $T_{1}$ of NV ranges from 6 ms at
room temperature \cite{D2} to $28\sim 265$ s at low temperature \cite{D3}.
In addition, later experimental progress \cite{D5} with isotopically pure
diamond has demonstrated a longer dephasing time to be $T_{2}$ $=1.8$ $ms$.
Therefore, the dissipation and decoherence of NV are negligible.

\section{Conclusion}

We have devised a concrete hybrid system to engineer photonic
MI-SF phase transition in a $2D$ square lattice of TLRs coupled to a single
NV encircled by a PCQ. We find that the interplay between the on-site
repulsion and the nonlocal tunneling leads to the photonic
localization-delocalization transition. In the presence of dissipation, the
phase boundary can be obtained by the mean-field approach and the quantum
jump technique. Facilitated by high levels of connectivity in circuit QED,
experiments combining both scalability and long coherence times are expected
in the coming few years, at that stage the investigation of photonic QPT
using TLR lattice systems can therefore be easier to realize.

\begin{acknowledgments}

We thank Xiaobo Zhu and Zhangqi Yin for enlightening discussions.
This work is supported partially by the National Research Foundation and
Ministry of Education, Singapore (Grant No. WBS: R-710-000-008-271), by the
National Fundamental Research Program of China under Grant No. 2012CB922102,
and by the NNSF of China under Grants No. 11274351 and No. 11204196.

\end{acknowledgments}


\begin{thebibliography}{99}
\bibitem{expl1} Here, the central coupler may be conceived as a Josephson
ring circuit \cite{Koch}, or a current-biased Josephson junction phase qubit
\cite{CBJJ}, or a capacitive coupling element \cite{Hu}, or an active
non-reciprocal devices as proposed in \cite{Kam}.
\end{thebibliography}

\end{document}